\documentclass[twocolumn,showpacs,preprintnumbers,amsmath,amssymb,prl]{revtex4}

\usepackage{graphicx}
\usepackage{dcolumn}
\usepackage{bm}

\begin{document}
\newcommand{\kp}{{\bf k$\cdot$p}\ }
\newcommand{\Pp}{{\bf P$\cdot$p}\ }

\preprint{APS/123-QED}

\title{Spatial control of electron spins by electric and magnetic fields in double
quantum wells}

\author{Pawel Pfeffer$\footnote{pfeff@ifpan.edu.pl}$ and Wlodek Zawadzki}
 \affiliation{Institute of Physics, Polish Academy of Sciences,\\
Al. Lotnikow 32/46, 02-668 Warsaw, Poland}

\date{\today}

\begin{abstract}
A system of two quantum wells (QW), one made of HgCdTe and the other of
HgCdMnTe, subjected to electric and magnetic fields $F$ and $B$ parallel to
the growth direction, is proposed and described theoretically. It is shown
that in such a system the spin $g$ factor of mobile electrons strongly
depends on the sign and magnitude of electric field. Adjusting $F$ at a
constant $B$ one can transfer almost all electrons into one or the other
QW and polarize their spins along the desired orientation. Changing $B$ at a
constant $F$ can produce a similar transfer and polarization effect.
Possible applications of the spatial reservoirs filled with spin-polarized
electrons are discussed.
\end{abstract}

\pacs{73.21.Fg$\;\;$73.61.Ga$\;\;$75.25.-j}

\maketitle

Spin systems in semiconductors have always been an important part of solid state investigations but in the last years they attracted great interest due to possible spintronic applications. In order to process quantum information one needs to manipulate fast and coherently the local
spins which could be done applying external magnetic fields. This, however, requires their
precise control at small length scales which is not easy. Salis et al.$^{1}$ demonstrated that one can control the spin splitting and
spin coherence using an external electric field employing parabolic Ga$_{1-x}$Al$_x$As quantum wells (QWs) and manipulating
the electron wave functions within such wells. The experiments of Salis et al were described by the present authors$^{2}$ using the {\kp} theory
for nonparabolic bands. The {\kp} theory was adapted to heterojunctions in which the effective mass and the spin
$g$ factor can depend not only on the energy but also on the spatial variable along the growth direction.

\begin{figure}
\includegraphics[scale=0.55,angle=0, bb = 250 20 202 300]{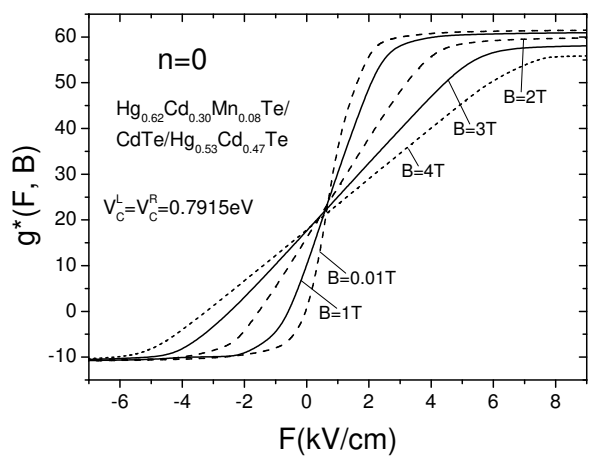}
\caption{\label{fig:epsart}{Calculated spin $g$ factor for $n = 0$ Landau level in two quantum wells of equal depths versus external electric field $F$ parallel
 to the growth direction for fixed magnetic fields $B$. It is seen that the $g$ factor sharply depends on the magnitude and sign of $F$. The widths of both wells are 9 nm, the barrier width is 4 nm.}} \label{fig1th}
\end{figure}

In the present work we propose and describe a system that will manipulate electron spins in more sophisticated ways that those realized in Ref. 1. Three
additional possibilities are envisaged. First, the electron spin will respond not only to the intensity but also to the sign of the electric field $F$.
Second, the spins will respond in a nonconventional way to the magnitude of magnetic field $B$. Third, it will be possible to create reservoirs
 of spin-polarized electrons in specific spatial regions. The proposed system consists of two rectangular semiconductor quantum wells separated by a barrier. The left well is made of HgCdMnTe alloy, the right one of HgCdTe alloy, the barriers are made of CdTe. Chemical compositions are chosen to make the well depths exactly or almost equal to each
 other. The alloy HgCdMnTe is a dilute magnetic (semimagnetic) material in which an external magnetic field creates an additional magnetization by ordering Mn ions.
This additional magnetization results in a higher spin splitting of electron energies characterized by a higher effective spin-splitting factor $g$. By
applying a positive or negative external electric field parallel to the growth direction one can transfer free electrons to either magnetic or nonmagnetic well
which will strongly affect their overall spin $g$ factor. This way, $g$ will depend not only on the electric field intensity but on its sign
as well. The other properties will become clear once the system is physically and mathematically described below.

Since the proposed alloys are relatively narrow-gap materials,
we use the three-level {\Pp} model (3 LM) for their band structures. In our description the energy gap and other band parameters are functions of $z$. The model explicitly takes into account 8 bands arising from the  $\Gamma^v_7$,  $\Gamma^v_8$ (double degenerate) and  $\Gamma^c_6$ levels at the center of the Brillouin
zone and treats the distant (upper and lower) levels as a perturbation. The resulting bands are spherical but nonparabolic. Our formulation includes external
magnetic and electric fields parallel to the growth direction. The multi-band {\Pp} theory, which is the {\kp} theory generalized for the presence of an external magnetic $\textbf{B}$ and electric $\textbf{F}$ fields, has the form$^2$
$$
\sum_l\left[\left( \frac{P^2}{2m_0}+E^{l}+{H}^{l}_{M}+V(z)+eFz-{\cal E})\right)\delta_{l'l} \right.
$$
\begin{equation}
\left.+\frac{1}{m_0}{\bf p}_{l'l}\cdot\bf{P} +\mu_{\rm B} {\bf B} \cdot{\bm \sigma}_{l'l} \right]f_l =0 \;\;,
\end{equation}
where ${\cal E}$ is the energy, ${\bf P} = {\bf p}+e{\bf A}$ is the kinetic momentum, ${\bf A}$ is the vector
potential of magnetic field $\textbf{B}$, and ${\bm \sigma}_{l'l} = (1/\Omega)\langle u_{l'}|{\bm \sigma}|u_l\rangle$. Here ${\bm
\sigma}$ are the Pauli spin matrices, $\Omega$ is the volume of the unit cell, $u_l$ are periodic amplitudes of
the Luttinger-Kohn functions, $\mu_B = e\hbar/2m_0$ is the Bohr magneton, ${\bf p}_{l'l}$ are the interband
matrix elements of momentum and ${H}^{l}_{M}$ is the exchange interaction between mobile electrons and those
localized around the Mn ions. The sum in Eq. (1) runs over
all bands $l = 1, 2,...,8$ included in the model, $l'=1, 2,...,8$ runs over the same bands, $E^{l}$ are the
band-edge energies, see below.
Within 3LM there exists the interband matrix element of momentum $P_0$ and that of the spin-orbit interaction $\Delta_0$ (see Ref. 3), as well as
two matrix elements related to magnetic Mn ions:  $\alpha$ and $\beta$, see Refs. 4 and 5. The latter
represent constants of the $s$ - $d$ and $p$ - $d$ exchange integrals related to the conduction band $|S\rangle$ and valence bands $|P\rangle$, respectively.
\begin{figure}
\includegraphics[scale=0.55,angle=0, bb = 250 20 202 340]{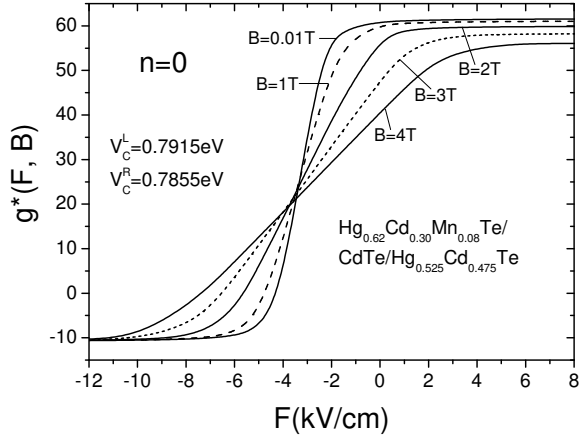}
\caption{\label{fig:epsart}{The same as in Fig. 1 for two QWs of slightly different depths. The overall picture is similar to that in Fig. 1 but the scale of electric fields is considerably shifted.}} \label{fig2th}
\end{figure}
For rectangular wells the potential $V(z)$ describes conduction band offsets and barriers. The valence offsets are automatically determined by corresponding energy gaps.

Equation (1) represents an 8$\times$8 system of equations for eight envelope functions $f_l(\textbf{r})$. Since we are interested in the eigenenergies
and eigenfunctions for the conduction band, we express the valence functions $f_3,...f_8$ by the conduction functions $f_1$ and $f_2$ for the spin-up and
spin-down states, respectively. The magnetic field $\textbf{B} || \textbf{z}$ is described by the asymmetric gauge $\textbf{A} = [-By, 0, 0]$ and
the resulting envelope functions have the general form $f_l = exp(ik_xx)\Phi_n[(y-y_0)/L]\chi_l(z)$, where $\Phi_n$ are the harmonic oscillator functions,
$y_0 = k_xL^2$ in which $L = (\hbar/eB)^{1/2}$ is the magnetic radius. After some manipulation the effective Hamiltonian for $f_1$
and $f_2$ functions is
\begin{equation}
\hat{H}=\left[ \begin{array}{cc}
\hat{A}^+&\hat{K}\\
\hat{K}^\dagger&\hat{A}^-
\end {array}
 \right]\;\;,
\end{equation}
where
$$
\hat{A}^{\pm}=V(z)+eFz \pm\frac{A_M}{2}-\frac{\hbar^2}{2}\frac{\partial}{\partial z} \frac{1}{m^*( {\cal E},
z)}\frac{\partial}{\partial z}
$$
\begin{equation}
+\frac{P^2_x+P^2_y}{2 m^*_I( {\cal E}, z)} \pm\frac{\mu_BB}{2}g^*( {\cal E}, z)
\;\;,
\end{equation}
in which the effective mass is
\begin{equation}
\frac{m_0}{m^*({\cal E}, z)}=1+C-\frac{1}{3}E_{P_0}\left(\frac{2}{\tilde E_0}+ \frac{1}{\tilde G_0}\right) \;\;,
\end{equation}
and the $g$-factor is
\begin{equation}
 g^*({\cal E}, z)=2+2C'+\frac{2}{3}E_{P_0}\left(\frac{1}{\tilde E_0}- \frac{1}{\tilde G_0}\right)+g^*_M(B, z)\;\;,
\end{equation}
where ${\tilde E_0} = E_0 - {\cal{E}}+V(z)+eFz$ and ${\tilde G_0} = G_0 - {\cal{E}}+V(z)+eFz$\;\;.
Here $G_0 = E_0 + \Delta_0$.  The spin splitting is related to the band structure [the last terms in Eq. (3)] and the semimagnetic interaction. Explicitly $g^*_{M}(B, z) = A_M/\mu_BB$, where $A_M$= x$\alpha \langle S_z\rangle$, x is the mole fraction of magnetic ions. The average spin component parallel to the applied magnetic field is $\langle S_z\rangle = -S_0B_S$(y), the total Mn spin is $S$ = 5/2 and $B_S$(y) is the Brillouin function. We take x = 0.08, $S_0$ = 1.02, $\alpha$ = 0.61 eV and $\beta$ = -0.62 eV$^6$. The small off-diagonal matrix elements $\hat{K}$ and $\hat{K}^\dagger$ in Eq. (2) appear due to inversion asymmetry of the system along the growth direction $z$, which results in an additional Bychkov-Rashba spin splitting, see below. The effective mass and the $g$ factor in Eqs.(4) and (5), respectively, are written neglecting small contributions of the semimagnetic $\beta$ terms in the valence bands.
\begin{figure}
\includegraphics[scale=0.55,angle=0, bb = 140 25 282 320]{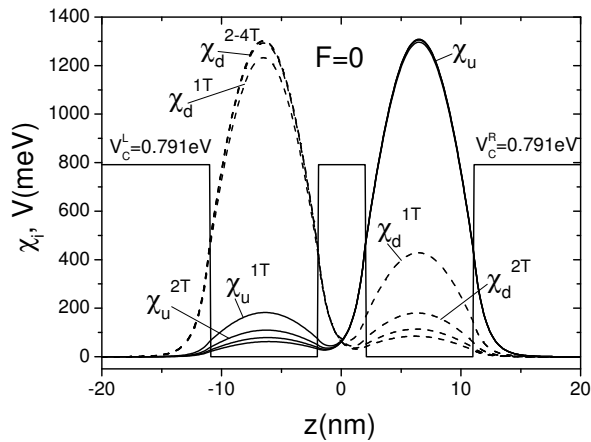}
\caption{\label{fig:epsart}{Calculated wave functions of $|0+\rangle$ and $|0-\rangle$ states for two QWs of equal depths at vanishing electric field $F$ and fixed magnetic fields $B$ (indicated by superscripts). Spin-up functions - solid lines, spin-down functions - dashed lines. Electrons in $|0-\rangle$ state are located mostly in left QW.
}} \label{fig3th}
\end{figure}
Equations $\hat{A}^+ f^+_n={\cal{E}}^+_n f^+_n$ and $\hat{A}^- f^-_n={\cal{E}}^-_n f^-_n$ are first solved separately for the system of two QWs at fixed values of the fields $B$ and $F$ using simple boundary conditions at the interfaces for the functions and their derivatives $\chi|_+ = \chi|_-$ and $[(1/m^*)\partial \chi/\partial z)]|_+=[(1/m^*)\partial \chi/\partial z)]|_-$. The transverse motion in $x$ - $y$ plane is quantized into the Landau
levels $\hbar\omega_c (n+1/2)$. The electric field term $eFz$ is small but not negligible compared to the offsets of QWs.

\begin{figure}
\includegraphics[scale=0.55,angle=0, bb = 140 25 282 320]{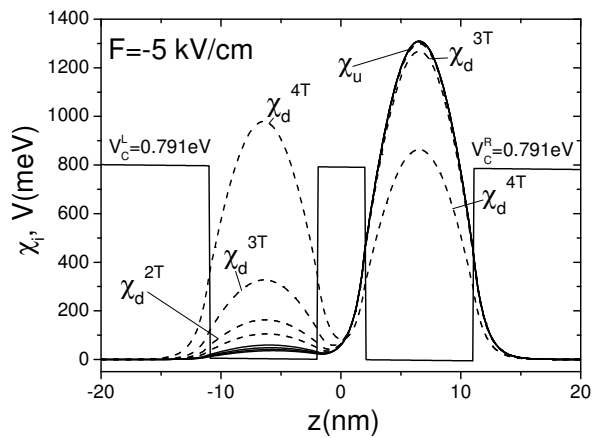}
\caption{\label{fig:epsart}{The same as in Fig. 3 but for electric field $F$ = -5 kV/cm. Electrons in the lowest $|0+\rangle$ state are located in right QW.}} \label{fig4th}
\end{figure}

We will consider two structures. In the first, the chemical compositions of QWs are such that without external fields their depths (conduction band offsets $V_C$) are identical and equal to 0.7915 eV. In the second, the the depth of the left HgCdMnTe QW is 0.7915 eV, while the depth of the right HgCdTe QW is 0.7855 eV. The zero of energy ${\cal E}$ is taken at the bottom of left QW.
\begin{table}
\caption{Band parameters of alloys for different chemical compositions $x$ and $y$, as used in the
calculations, $C$ and $C'$ are far-band
contributions to the band-edge values. Interband matrix element of momentum $P_0$ is taken to be independent of $x$ and $y$: $E_{P_0}=2\hbar^2P^2_0/m_0$ = 18 eV.}

\begin{ruledtabular}

\begin{tabular}{ccccc}
&CdTe&Hg$_{.62}$Cd$_{.3}$Mn$_{.08}$Te&Hg$_{.53}$Cd$_{.47}$Te&Hg$_{.525}$Cd$_{.475}$Te\\
\hline
 E$_0$(eV) & -1.606 & -0.495 & -0.5125& -0.5211 \\
G$_0$(eV)&-2.516&-1.510&-1.5126&-1.5203\\
C&-0.104&-0.104&-0.104&-0.104\\
C'&-0.479&-0.479&-0.479& -0.479\\
V$_C$(eV)&---&0.7915&0.7915& 0.7855\\
m$^*_0$/m$_0$&0.093&0.0343&0.0353&0.0359\\
g$^*_0$&-1.66&-15.25&-14.437&-14.093\\
\end{tabular}
\end{ruledtabular}
\end{table}
In Figures 1 and 2 we show calculated electron spin $g$ values for the two structures versus an external electric field $F$ at fixed magnetic field intensities $B$. It is seen that, indeed, the $g$ values strongly depend not only on the value of $F$ but also on its sign. Generally speaking, if the positive electric field pushes the electrons into semimagnetic QW they feel the additional magnetization of Mn ions and the overall $g$ value is higher. It is also seen that, by changing chemical compositions of the alloy and the corresponding depth of right QW (without Mn ions), one strongly shifts the scale of electric fields, while the overall pattern of $g$ behavior and its dependence on $B$ remains similar. Since we assume that the number of mobile electrons is small and the temperature is low, so the electrons will occupy the lowest Landau and spin level, one can use Figs. 1 and 2 to decide which spin level will be occupied for given values of $F$ and $B$. As far as the wavefunctions in the system of two QWs are concerned, there are two factors that determine where the wave function (WF) is predominantly located. First, if the electron masses are the same for both wells, the WF is located mostly in the deeper QW. Second, if both QWs have the same depth, WF is located mostly in QW with the higher electron mass.

\begin{figure}
\includegraphics[scale=0.55,angle=0, bb = 140 25 282 320]{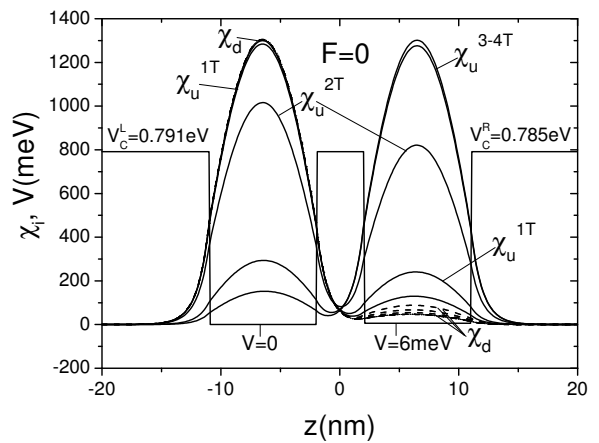}
\caption{\label{fig:epsart}{The same as in Fig. 3 but for two QWs of different depths at $F$ = 0. Electrons in the lowest $|0-\rangle$ state are located in left QW.
}} \label{fig5th}
\end{figure}

We consider first the case of two QWs of equal depths. Taking $F$ = 0 and 2T$\le B \le$4T, it follows from Fig. 1 that the spin factor $g$ is positive, so electrons will occupy the lowest level $|0-\rangle$. Calculated WF for this case is shown in Fig. 3 and it is seen that WF is almost entirely concentrated in left QW, which means that the electrons will be located there and they will have the spin-down orientation. If now one applies $F$ = -5kV/cm and 1T$\le B \le$4T, it is seen from Fig. 1 that at these fields $g$ is negative and the electrons will occupy the $|0+\rangle$ state. Calculation for this situation is shown in Fig. 4 and it is seen that WF for $|0+\rangle$ state is strongly concentrated in right QW. Thus, the application of electric field $F$ at constant $B$ transfers electrons from left to right QW and changes their spin orientation from spin-down to spin-up.

A similar effect can be obtained in the system of QWs having different depths. Taking $F$ = 0 and 1T$\le B \le$4T it is seen in Fig. 2 that in this regime the spin $g$ value is positive, so the occupied state is $|0-\rangle$. It follows from Fig. 5 that the calculated WF for the spin-down state is in left QW and the electrons will be located there. If now one applies the fields $F$ = - 5kV/cm and $B$ = 1T, the $g$ value is negative (see Fig. 2) and the occupied state is $|0+\rangle$. The calculated WF for this situation is shown in Fig. 6, one can again see that WF for this state is almost completely concentrated in right QW. Thus, by changing the electric field one can move electrons from left to right QW and change their spin orientation from spin-down to spin-up.

\begin{figure}
\includegraphics[scale=0.55,angle=0, bb = 140 25 282 320]{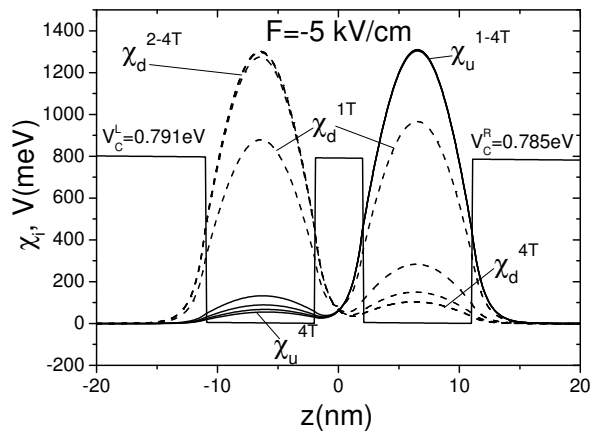}
\caption{\label{fig:epsart}{The same as in Fig. 5, but for $F$ = -5kV/cm. At small magnetic fields $B$ the electrons are in $|0+\rangle$ state and they are located in right QW, while for high $B$ they are in $|0-\rangle$ state and are located in left QW, see text.}} \label{fig6th}
\end{figure}

Finally, we consider the case of unequal QWs taking $F$ = -5kV/cm at different magnetic fields. For $B$ = 1T one reads from Fig. 2 that $g \approx$ -4.6, so the electrons will be in the $|0+\rangle$ state and from Fig. 6 one concludes that they will be concentrated in right QW. If now for the same $F$ =-5kV/cm the magnetic field is changed to 2T$\le B \le$4T, it follows from Fig. 2 that $g$ becomes positive, so the occupied state is $|0-\rangle$. The wave function for this state at the above fields is strongly concentrated in left QW, see Fig. 6. Thus changing the intensity of $B$ alone at a constant electric field $F$ one will reverse the electron spin and transfer electrons from right to left QW. In other words, electrons will move \emph{along} the magnetic field, which is an unusual effect.

As mentioned above, an asymmetric structure of two QWs with a nonvanishing electric field applied along the growth direction $z$ will result in an additional Bychkov - Rashba spin splitting and spin mixing due to structure inversion asymmetry (SIA)$^{7, 8}$. We considered this effect using Eq. (2). The SIA terms couple $|0+\rangle$ and $|1-\rangle$ states. It turns out that the additional spin splitting is quite small and an admixture of opposite spin constitutes not more than 7.6 $\%$ of the the total wave function squared at electric fields of our interest. It should be mentioned that the same mechanism was used by Nowack et al$^{9}$
to control a single-electron spin in a quantum dot by an electric field.

  Spin polarized electrons were created in the past by optical orientation
(OO), exciting III-V semiconducting compounds (mostly GaAs) with
circularly polarized light. Fabricated sources were subsequently used
for experiments in various domains of physics: electron collisions with
atoms, surface magnetism, parity non-conservation in inelastic
scattering at high energies, etc$^{10}$. The system we propose offers
important advantages in comparison with OO. First, it creates almost
completely spin-polarized  electron gas, while OO gives 40-50$\%$
efficiency. Second, due to spin splitting by magnetic field, electrons
stay a long time in their spin state, while in OO regime they quickly
recombine. Third, the spin-polarized electrons are in spatial reservoirs
which should facilitate applications. Finally, our system offers the
following interesting possibility. If the temperature is raised, so that
both $|0-\rangle$ and $|0+\rangle$ levels are populated, it follows from Figs. 3 and 6 that, for 3T$\le B \le$4T, the
spin-down electrons will be in left QW while the spin-up electrons will
be in right QW. Thus the system spatially separates electrons with
opposite spins which amounts to an effective Stern-Gerlach experiment.
The latter is, as is well known, not possible with electrons in a
vacuum, see Ref. 11.

\end{document}